\documentclass[aps,pra,twocolumn,showpacs,superscriptaddress,groupedaddress]{revtex4-1}
\usepackage{graphicx}    % needed for figures
\usepackage{dcolumn}    % needed for some tables
\usepackage{bm}             % for math
\usepackage{amssymb}   % for math
\usepackage{amsmath}    % for multiline equation
\usepackage{mathrsfs}    % for mathscr{}
\usepackage{commath}   % for absolute value
\usepackage{subfigure}    % for subfigure
\usepackage{braket}
\usepackage{natbib}
\usepackage{diagbox}
\usepackage{float}
\begin{document}

\title{Quantum calculation of feedback cooling a laser levitated nanoparticle in the shot-noise-dominant regime}
\author{Changchun Zhong$^1$}
\email{zchangch@purdue.edu}
\author{Tongcang Li$^{1,2,3}$}
\author{F. Robicheaux$^{1,2}$}
\email{robichf@purdue.edu} 
\affiliation{$^1$Department of Physics and Astronomy, Purdue University, West Lafayette IN, 47907 USA}\affiliation{ $^2$Purdue Quantum Center, Purdue University, West Lafayette, IN, 47907 USA}
\affiliation{$^3$School of Electrical and Computer Engineering, Purdue University, West Lafayette, IN, 47907 USA}

\date{\today}

\begin{abstract}

In this paper, results of quantum calculations are presented for feedback cooling of an optically trapped nanoparticle in the laser-shot-noise-dominant regime. We numerically investigate the system using both parametric and force feedback cooling schemes. For the same measurement efficiency, the cooling limit from the force feedback is lower than that from the parametric feedback. We also develop a set of semi-classical equations for feedback cooling that accurately match the quantum results. It is demonstrated, by rescaling the semi-classical equations, that the cooling dynamics is uniquely determined by the parameter set: the feedback strength, the measurement efficiency and the change of occupation number over one oscillation period due to the shot noise. The minimum occupation number is determined by the measurement efficiency and the change of occupation number over one period.

\end{abstract}

%\pacs{42.50.Wk, 07.10.Pz, 62.25.Fg}
\maketitle

\section{introduction}

The system of optically trapped nanoparticles has recently emerged as an exciting candidate for tests of quantum mechanics at the mesoscale \cite{RTCC,SPFC,QMOC,CRAT,COUA}. It helps not only in our understanding of quantum fundamentals, such as the role decoherence plays in the quantum-classical transition, but also in the development of practical applications, such as ultrasensitive metrologies \cite{FRCO,NTMU,SFDU,NOWO,QSOM,DCRW,COUA}. Because the nanoparticle is levitating, it is not directly attached to any mechanical device, leading to good thermal isolation. In Ref. \cite{DMOP}, it is reported that the photon shot noise overwhelms the thermal noise by at least a factor of $25$ when the particle is trapped in ultrahigh vacuum (the pressure is about $10^{-8}$ mbar). Thus, the shot noise from the trapping laser becomes the particle's major source of decoherence. In this paper, we present results on the quantum feedback cooling of a levitated nanoparticle in this shot-noise-dominant regime. The feedback signal is obtained through continuously measuring the particle's position. Due to the measurement back-action, the system state evolves stochastically, which is described by a stochastic master equation or equivalently by a stochastic Schr$\ddot{\text{o}}$dinger equation \cite{ASMO,ASIT}. The measured position is used to modify the system Hamiltonian such that cooling of the center of mass degrees of freedom is achieved \cite{FCOQ}.

The force feedback and the parametric feedback schemes are widely used in cooling an optically trapped nanoparticle \cite{MCOA,SPFC}. They are realized by tuning the force exerted on the particle (force feedback) or by changing the trapping laser intensity (parametric feedback). In this paper, the cooling of an optically trapped nanoparticle is simulated using these two feedback cooling schemes. For each cooling scheme, we calculate and compare the steady state occupation number as a function of the feedback strength and the measurement efficiency. It is demonstrated that a lower cooling limit can be reached by force feedback cooling than by parametric feedback cooling for the same measurement efficiency. Since the quantum calculations are time consuming, we develop a set of semi-classical equations for modeling the feedback cooling, where a concept of classical measurement uncertainty is introduced \cite{SNDR}. It is shown that the quantum and semi-classical results of the cooling limit are the same. Remarkably, by rescaling the semi-classical equations, we find that the optimal cooling limit is uniquely determined by the parameter set: $\Delta n$ (defined as the change of occupation number in a vibrational period due to the shot noise), the feedback cooling strength, and the measurement efficiency. Our study provides a useful guide and framework for the community to think about how those parameters affect the the feedback cooling of levitated nanoparticles.

The measurement efficiency plays a significant role in both the force feedback and the parametric feedback cooling. In order to achieve ground state cooling ($\braket{n}<1$), a suitable measurement efficiency must be reached. For the force feedback scheme, more than ten percent measurement efficiency is sufficient for cooling the nanoparticle to the ground state, while a higher efficiency (more than forty percent) is needed for the parametric feedback to achieve ground state cooling. In practice, the measurement efficiency is determined by the photon collection efficiency and the actual measurement scheme. For a given experimental setup, the measurement efficiency should have an upper bound if we assume a perfect photon detection efficiency. To evaluate this upper bound is important by itself and is significant in guiding cooling experiments. Derivation of the best achievable efficiency for a given measurement technique is beyond the scope of this paper. 

The paper is organized as follows. Section \ref{s1} introduces the system we consider. In Sec. \ref{s2}, the theory of continuous quantum measurement is briefly reviewed and two feedback cooling schemes are introduced. In Sec. \ref{s3}, the quantum and semi-classical results of cooling by the force feedback and the parametric feedback cooling methods are discussed. Finally, a summary and conclusion are given in Sec. \ref{s4}. SI units are used throughout the paper.

\section{The laser levitated nanoparticle} \label{s1}

%-------
\begin{figure}
\includegraphics[width=6.0cm,height=4.0cm]{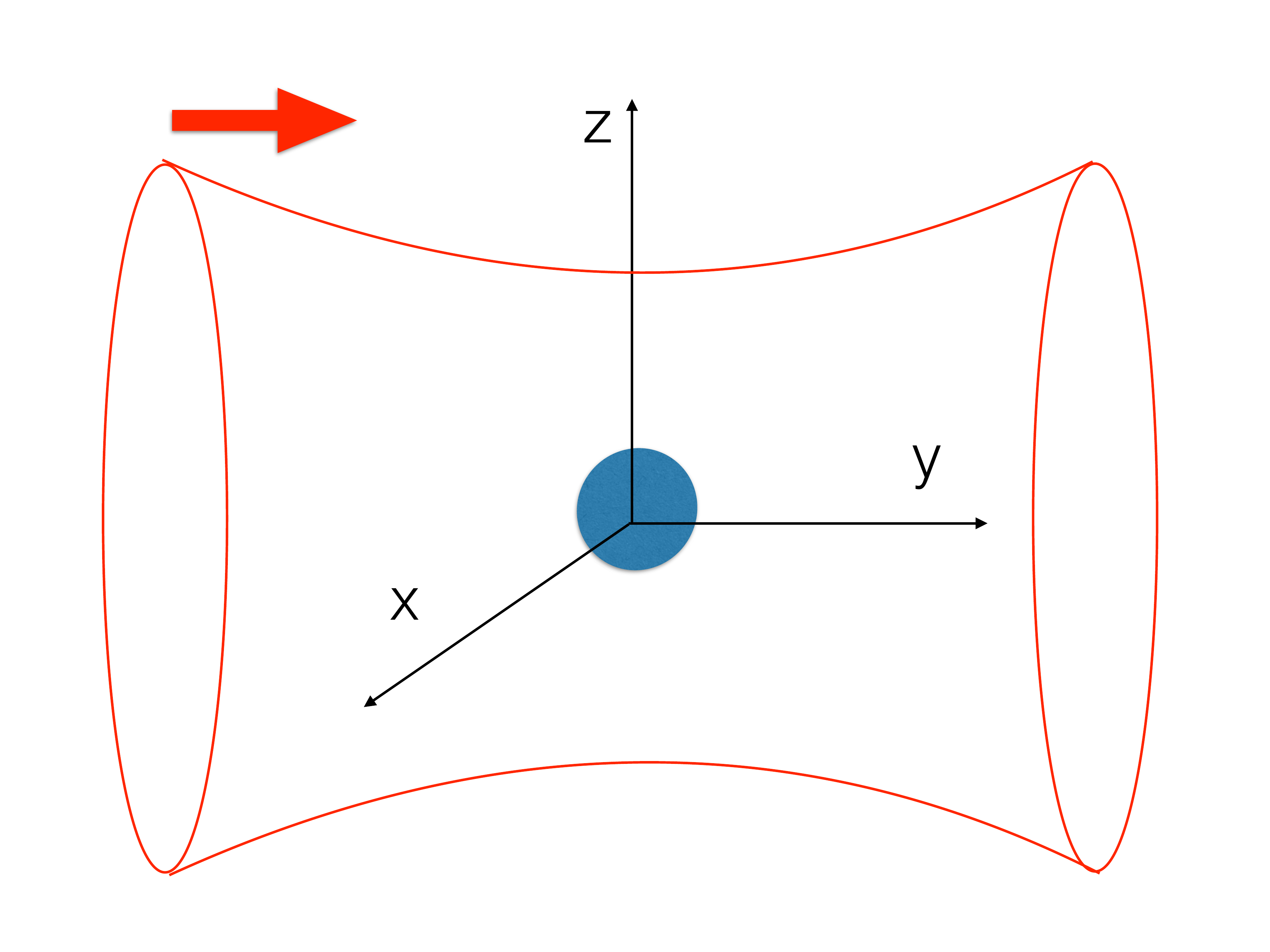}
\caption{A nanoparticle with mass $m$ is trapped at the focus of a laser beam (schematically shown by the red line), which is polarized in the $z$ direction and propagating in the positive $y$ direction (shown by the red arrow). \label{5fg1}}
\end{figure}
%-------

A laser levitated nanoparticle in ultrahigh vacuum is well isolated from its thermal environment. Due to the isolation, the trapping laser is the particle's major source of heating \cite{DMOP}, which results from the recoil of randomly scattered photons. In this paper, we consider a nanoparticle trapped in the focus of a linearly polarized laser beam, as shown in Fig. (\ref{5fg1}). The laser field is polarized in $z$ and propagating in the positive $y$ direction. Using a scattering model \cite{DATA,CDRE,DTMP,DORD}, one can show that the system density operator $\bm{\rho}$ follows the master equation
\begin{equation}\label{5e1}
\frac{d\bm{\rho}}{dt}=\frac{1}{i\hbar}[\bm{H},\bm{\rho}]-\kappa[\bm{x},[\bm{x},\bm{\rho}]],
\end{equation}
where the wavelength of the light is assumed to be much larger than the size of the nanoparticle. $\bm{H}$ is the system Hamiltonian which will be given below. The last term is from the decoherence due to photon scattering which localizes the position of the nanoparticle. $\bm{x}$ is the position operator and $\kappa=\dot{E} m/\hbar^2$ is the interaction strength, where $\dot{E}$ is the shot noise heating from the trapping laser and $m$ is the particle mass. $\dot{E}$ takes the form \cite{SPFC,DMOP}
\begin{equation}\label{5e2}
\dot{E}=\frac{8\pi J_p}{3}\left(\frac{k^2_0}{4\pi\epsilon_0}\right)^2\alpha^2\frac{\hbar^2k^2_0}{2m},
\end{equation} 
where $J_p$ is the laser photon flux, $k_0$ is the incoming wave vector, $\alpha=4\pi\epsilon_0R^3(\epsilon-1)/(\epsilon+2)$ is the particle polarizability, $\epsilon_0$ is the vacuum dielectric constant and $\epsilon$ is the relative dielectric constant. It is worth mentioning that, because of the detailed pattern of dipole radiation, the shot noise is different in each degree of freedom by the factors $\zeta_x=\zeta_y=2/5$, $\zeta_z=1/5$. The factors must be added to the corresponding degree of freedom when evaluating the shot noise \cite{SPFC}.

\begin{table}[htp]
\caption{The parameters for diamond and fused silica with sizes $R\simeq50$ nm trapped in a laser beam. The laser has wavelength $\lambda=1064\text{ nm}$, power $P=70\text{ mW}$ and an objective lens with numerical aperture NA$=0.9$. The parameters are given in the $x$ degree of freedom. }\label{tab1}
\begin{center}
\begin{tabular}{|c|c|c|c|c|c|c|c|c|c|}
\hline
\text{ }  & $\epsilon$  & mass (kg)  & $\omega_x$ (kHz)  & $\dot{E}$ (W)  & $\Delta n$  \\
\hline
diamond &  $5.7$  & $1.79\times10^{-18}$   &   $2\pi\times 454$   &  $4.55\times10^{-24}$   &   0.033  \\
\hline
silica      &  $2.1$   & $1.13\times10^{-18}$   &   $2\pi\times 374$   &  $1.32\times10^{-24}$   &   0.014  \\
\hline
\end{tabular}
\end{center}
\label{default}
\end{table}

The nanoparticle is trapped at the focus by an optical gradient force. If one models the laser beam as Gaussian and takes a small oscillation approximation \cite{LAWT,OAAC}, each degree of freedom of the particle oscillates as an independent harmonic oscillator along its principal axis. The Hamiltonian can be written as
\begin{equation}
\bm{H}=\sum_{i=x,y,z}\hbar\omega_i(\bm{a}_i^\dagger\bm{a}_i+\frac{1}{2}),
\end{equation}
where $\bm{a}_i^\dagger,\bm{a}_i$ are creation and annihilation operators respectively. The oscillation frequencies $\omega_i$ were shown to be \cite{LAWT,OAAC}
\begin{equation}\label{5e44}
\begin{split}
\omega_{x,z}&\simeq\sqrt{\frac{\alpha}{m}}\frac{E_0}{w_0},\\
\omega_y&\simeq\sqrt{\frac{\alpha}{2m}}\frac{E_0}{y_0},
\end{split}
\end{equation}
where $y_0={\pi w_0^2}/{\lambda}$, $w_0=\lambda/(\pi \text{NA})$ is the beam waist, and $E_0$ is the field strength at the center of the laser focus. In this paper, we use a laser with wavelength $\lambda=1064$ nm, power $P_0=70$ mW, and an objective lens with numerical aperture NA$=0.9$. The particle size is on the order of $R=50\text{ nm}$. For later discussions, we evaluate the parameters related to feedback cooling and put them in Tab. \ref{tab1}. Unless specified otherwise, all numerical simulations in the following sections are performed using the parameters discussed above. Since each degree of freedom of the nanoparticle is essentially decoupled, we focus on only the $x$ degree of freedom in the following discussions. The notations without subindex all correspond to quantities in the $x$ degree of freedom.  

\section{Continuous quantum measurement and the feedback cooling scheme} \label{s2}

In quantum feedback control, a system needs to be continuously measured and the measured information is collected to modify the system Hamiltonian so as to achieve a targeted outcome \cite{FCOQ}. The system evolution conditioned on the measurement result can be derived by the theory of continuous quantum measurement \cite{ASIT}. For demonstration, we briefly summarize the theory and one may refer to Ref. \cite{ASIT} for a detailed introduction. 

Due to the back-action of a position measurement, the system state evolves stochastically according to
\begin{equation}\label{5e3}
\begin{split}
d\bm{\rho}=&\frac{1}{i\hbar}[\bm{H},\bm{\rho}]dt-\kappa[\bm{x},[\bm{x},\bm{\rho}]]dt\\
&+\sqrt{2\eta\kappa}(\bm{x\rho}+\bm{\rho x}-2\braket{\bm{x}}\bm{\rho})\sqrt{dt}dW,\\
x_i=&\braket{\bm{x}}+\frac{dW}{\sqrt{8\eta \kappa dt} },
\end{split}
\end{equation}
which is known as the stochastic master equation (SME) \cite{ASIT}. $x_i$ is the directly measured value of position. In our calculation, $x_i$ is time averaged to get a better estimate of the particle position $x_m$ (shown in Appendix \ref{a1}). $\braket{\bm{x}}=\text{tr}(\bm{\rho} \bm{x})$. The parameter $\eta$ is the measurement efficiency, which determines the uncertainty in the measured position, $\Delta x=1/\sqrt{8\eta\kappa dt}$. From Eq. (\ref{5e2}), a random momentum kick can be obtained $\Delta p=\sqrt{2\kappa dt}\hbar$. Thus, one immediately gets 
\begin{equation}\label{5e33}
\Delta x\Delta p=\frac{1}{\sqrt{\eta}}\frac{\hbar}{2},
\end{equation}
where the measurement efficiency $\eta$ is by definition smaller than one $\eta\leq 1$, and $\eta=1$ corresponds to the minimal uncertainty allowed by quantum mechanics. $dW$ is a standard normally distributed Gaussian random variable. Since $dW$ is random, there would be many solutions to the above equation and each realization $\bm{\rho}(t)$ defines a quantum trajectory. Equation (\ref{5e3}), excluding the last stochastic term, is the same as Eq. (\ref{5e1}). Actually, if all the measured information were lost, one would need to average all the possible quantum trajectories, which leads to Eq. (\ref{5e1}) due to the zero mean of $dW$. The evolution of the system can also be written in terms of a wave function $\ket{\psi}$,
\begin{equation}\label{5e4}
\begin{split}
d\ket{\psi}=&\{\frac{1}{i\hbar}\bm{H}dt-\kappa(\bm{x}-\braket{\bm{x}})^2dt\\
&+\sqrt{2\kappa}(\bm{x}-\braket{\bm{x}})\sqrt{dt}dW \}\ket{\psi},
\end{split}
\end{equation}
which is referred to as the stochastic Schr$\ddot{\text{o}}$dinger equation (SSE). In quantum simulations, the SSE is generally favored since the numerical calculation cost is much less than that required by the SME \cite{WFMC}.

The stochastic equation is conditioned on the measured position, which we can use to modify the system Hamiltonian. In this paper, we investigate two different feedback cooling schemes. The first one is force feedback. The force feedback was first used in cooling an optically trapped microsphere \cite{MCOA}. It works by exerting a force on the particle with the force direction opposite to the particle's instantaneous momentum. Thus, the modified Hamiltonian can be written as
\begin{equation}
\bm{H}=\hbar\omega(\bm{a}^\dagger\bm{a}+\frac{1}{2})+\gamma p_m\bm{x},
\end{equation}
where $p_m=m\dot{x}_m$ is the feedback momentum. $p_m$ can be obtained from the measured position $x_m$ according to the formula 
\begin{equation}\label{5e9}
\dot{x}_m\simeq -\omega x_m(t-T/4),
\end{equation}
where the small oscillation approximation is used and $T$ is the oscillation period. The parameter $\gamma$ is the force feedback strength, and $\bm{x}=\sqrt{{\hbar}/({2m\omega})}(\bm{a}^\dagger+\bm{a})$ is the position operator. In the next section, this Hamiltonian is used in the force feedback cooling by numerically calculating both SSE and SME. It is worth mentioning that an extra noise might be introduced depending on the way force feedback is implemented, such as force feedback by implementing radiation pressure \cite{MCOA}. This noise is another source of heating, which is not taken into account in the current discussion. 

The other cooling method is parametric feedback, where a single laser beam is used for both trapping and cooling \cite{SPFC,ANPF}. In the parametric feedback scheme, a signal at twice the oscillation frequency is obtained by multiplying the measured particle's position with its first time derivative $x_m(t)\dot{x}_m(t)$. This signal is then used to modify the laser trapping depth, which on average acts as a drag on the particle. The modified Hamiltonian can be written as
\begin{equation}
\bm{H}=\hbar\omega(\bm{a}^\dagger\bm{a}+\frac{1}{2})+\frac{\chi}{2}m\omega^2x_m\dot{x}_m\bm{x}^2,
\end{equation}
where $\chi$ is the parametric feedback strength ($\eta$ was used in our previous paper \cite{SNDR}). In the following section, this Hamiltonian is used in the SSE to simulate the quantum parametric feedback cooling.

\section{The numerical simulation of feedback cooling}\label{s3}

\subsection{Cooling by force feedback}

In this subsection, we present the numerical calculations of the force feedback cooling. Both SME and SSE are numerically solved. We first define an average occupation number of the nanoparticle as $\braket{n}=\braket{\bm{a}^\dagger\bm{a}}=\text{tr}(\bm{\rho}\bm{a}^\dagger\bm{a})$. In an experiment, it is this number that one wants to decrease to lower than one (the ground state). The particle occupation number $\braket{n}$ is calculated with respect to different values of the force feedback strength $\gamma$. Besides the quantum calculations, the cooling process is also simulated semi-classically and we will show later that the quantum calculations match the semi-classical results. The semi-classical occupation is defined as $\braket{n}=\braket{E}/\hbar\omega$, where the energy is calculated through $E=\frac{p^2}{2m}+\frac{1}{2}m\omega^2x^2-\frac{1}{2}\hbar\omega$. The semi-classical equation of motion for force feedback is given by
\begin{equation}
\begin{split}
m\frac{d^2x}{dt^2}&=-m\omega^2x-\gamma p_m,\\
x_i&=x+dW\cdot\Delta x,
\end{split}
\end{equation}
where $x_i$ is the directly measured position, $p_m$ is the feedback momentum which is obtained from the measured position as discussed in the previous section, and $\Delta x=\hbar/\sqrt{8\eta\dot{E} dt\cdot m}$. As time proceeds, the shot noise induces a random momentum kick on the particle by
\begin{equation}
p(t+dt)=p(t)+dW\cdot\Delta p,
\end{equation}
where $\Delta p=\sqrt{2\dot{E} dt\cdot m }$. It is worth mentioning that classically there is no theoretical limit to measure the position accurately. The uncertainty in the measured position $x_i$ is added to make it quantitatively satisfy Eq. (\ref{5e33}). Thus, the classical uncertainty in position and momentum also satisfies $\Delta x\Delta p=\frac{1}{\sqrt{\eta}}\frac{\hbar}{2}$ \cite{SNDR}. The classical uncertainty is fundamentally different from the quantum uncertainty, which intrinsically limits what we can know about physical observables.

\begin{figure}
\centering
\subfigure
{
\put(30,32){(a) Diamond}
\begin{minipage}[]{0.24\textwidth}
\includegraphics[width=4.2cm,height=3.2cm]{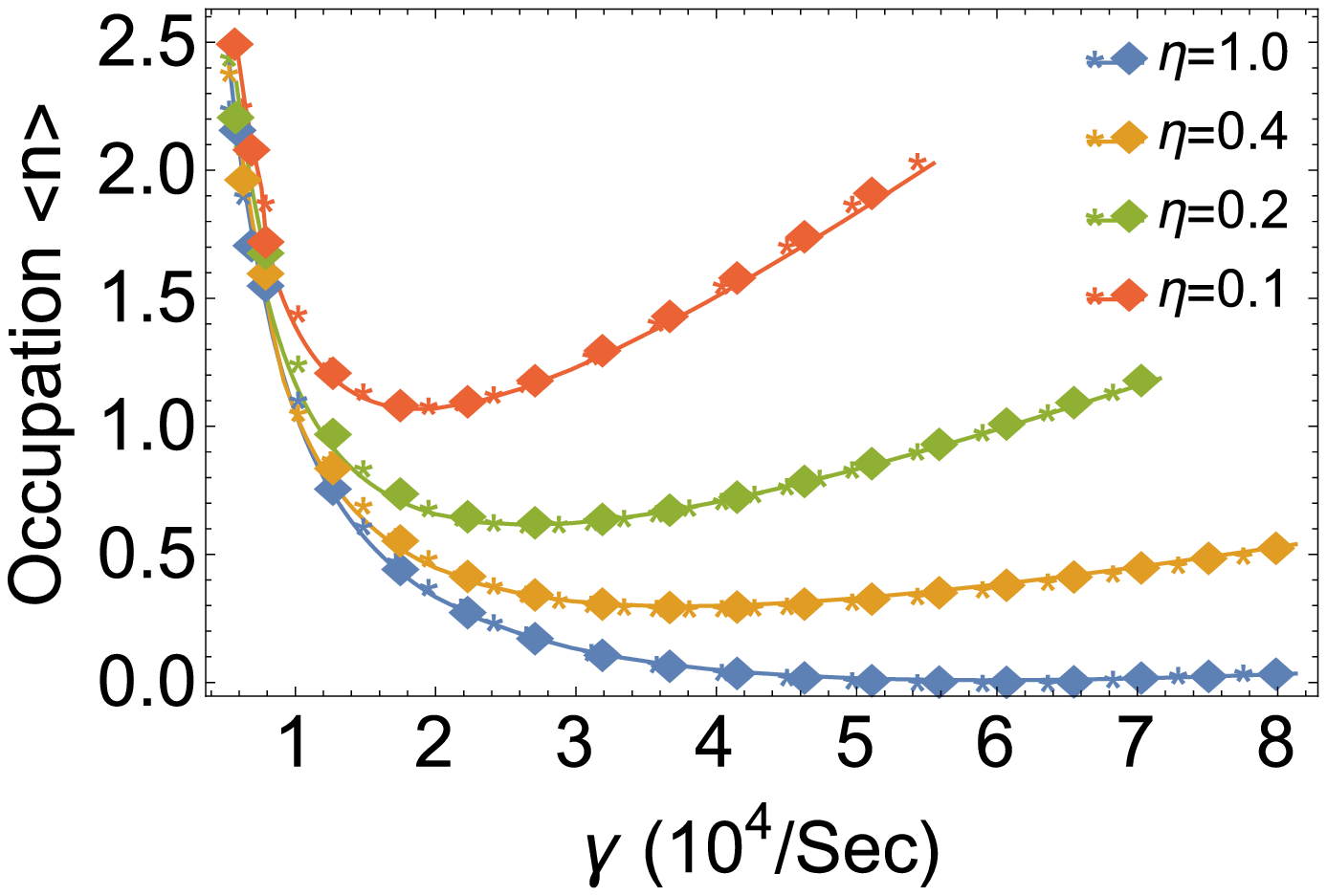}
\end{minipage}
\label{fadd1}
}
\subfigure
{
\put(25,32){(b) Silica}
\begin{minipage}[]{0.22\textwidth}
\includegraphics[width=4.2cm,height=3.2cm]{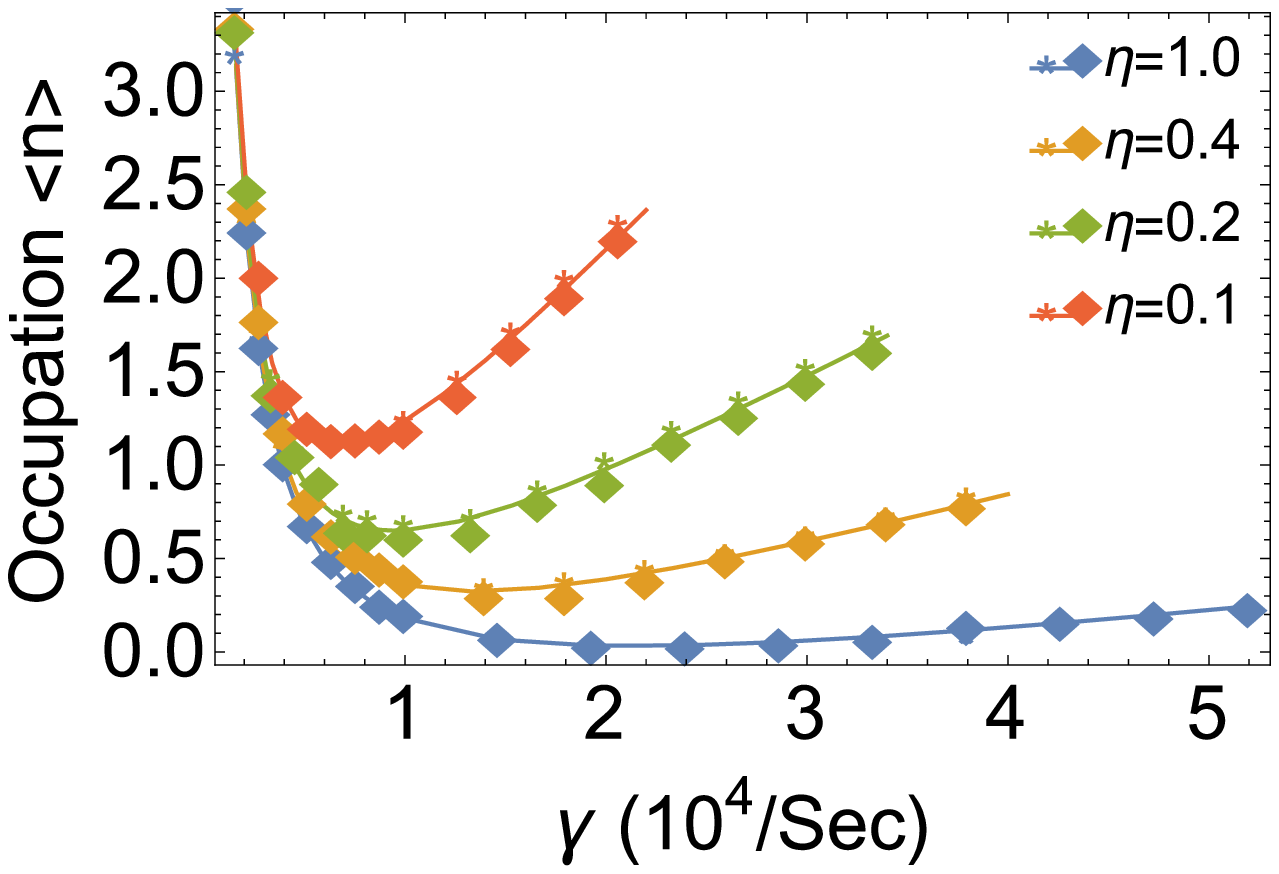}
\end{minipage}
\label{fadd2}
}
\caption{{(Color online)}  The steady state occupation in terms of the force feedback strength for diamond (left) and silica (right). The parameters in Tab. \ref{tab1} are used. The solid lines, the asterisks and the diamonds correspond to the SSE, SME and the semi-classical results respectively. The data shown by the colors red, green, yellow and blue are for four different measurement efficiencies $\eta=(1.0,0.4,0.2,0.1)$. \label{5fg2} } 
\end{figure}
%------------------

%-------
\begin{figure}
\includegraphics[width=8.0cm,height=5.5cm]{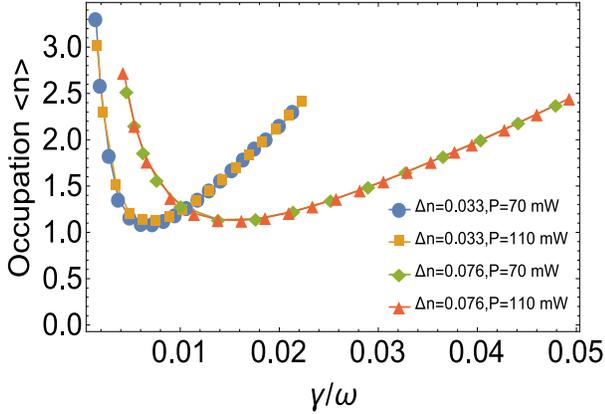}
\caption{{(Color online)} The steady state occupation in terms of the scaled force feedback strength for diamond, with the parameters given in Tab. \ref{tab1}. $\Delta n$ is tuned by changing the beam waist. The measurement efficiency is fixed at $\eta=0.1$. The solid lines are results from SSE and the symbols from semi-classical calculations. \label{5fg3}}
\end{figure}
%-------

The SSE and SME are numerically solved in the harmonic eigen-basis, and the detail of the numerical method is given in the appendix \ref{a1}. The semi-classical equations of motion are numerically solved using a fourth-order Runge-Kutta algorithm and the momentum kick is added to the nanoparticle at each time step. All simulations are repeated over one thousand times and data is collected by averaging over the different runs to reduce the random noise. 

The results are given in Figs. (\ref{5fg2}), which show the steady state occupation number with respect to the feedback strength for both diamond and silica. First, the quantum and semi-classical results match, even for the measurement efficiency $\eta=1.0$, where a variation is intuitively expected between semi-classical and quantum calculations. This match justifies the use of semi-classical equations in the further analysis, which are more intuitively revealing and less computationally demanding. Second, as the measurement efficiency increases, the steady state occupation number is smaller for the same feedback strength. A better measurement efficiency indicates a more accurate measured position, which in turn leads to better feedback cooling. Third, for a fixed measurement efficiency $\eta$, the steady state occupation number has a minimal point (the optimal cooling limit), which can be reached as the feedback strength is tuned. As one increases the feedback strength, the feedback cooling is strengthened, but the feedback procedure itself adds heat into the system due to the noise of the measured position. The competition between these cooling and heating leads to the curved structure, which has a minimal point. Fourth, even for the $\eta=0.1$ measurement efficiency, a steady state occupation number close to $\braket{n}=1$ can be reached, indicating the possibility of ground state cooling using the force feedback cooling scheme. Lastly, one can see that the optimal cooling limits are quite close for both silica and diamond when the measurement efficiency is chosen the same. We show below that the optimal cooling limit mainly depends on the measurement efficiency, but weakly depends on other parameters.

The fact that the quantum and semi-classical results match encourages us to further study the semi-classical equations, since they are intuitively revealing and can be solved more rapidly. We rescale the above semi-classical equations
\begin{equation}
\begin{split}
\frac{d^2\tilde{x}}{d\tilde{t}^2}&=-\tilde{x}-\tilde{\gamma}\tilde{p}_m,\\
\tilde{p}(t+dt)&=\tilde{p}(t)+dW\cdot\Delta\tilde{p},\\
\tilde{x}_i&=\tilde{x}+dW\cdot\Delta\tilde{x},
\end{split}
\end{equation}
where the scaled position $\tilde{x}=x/a_0$ with $a_0=\sqrt{\hbar/(2m\omega)}$, $\tilde{t}=t\omega$, $\tilde{p}=p/(m\omega a_0)$ and the scaled feedback strength $\tilde{\gamma}={\gamma}/{\omega}$. $\Delta\tilde{p}=\sqrt{{2\tilde{\dot{E}}}\cdot{d\tilde{t}}}$ and $\Delta\tilde{x}=\sqrt{1/(2\eta\tilde{\dot{E}} d\tilde{t})}$ with $\tilde{\dot{E}}=2\dot{E}/(\hbar\omega^2)$. We define an important (shown later) quantity 
\begin{equation}\label{5e13}
\Delta n\equiv \frac{2\pi\dot{E}}{\hbar\omega^2}=\pi\tilde{\dot{E}},
\end{equation}
which denotes the change in occupation number over one oscillation period. A detailed discussion of this quantity can be found in Ref. \cite{SNDR}. As shown in the scaled semi-classical equation, the dynamics of force feedback cooling is totally governed by the parameter set \{$\eta$,$\tilde{\gamma}$,$\Delta n$\}. In an experiment, if the measurement efficiency $\eta$ is fixed, and we assume the feedback strength $\tilde{\gamma}$ can be freely tuned, then the choice of $\Delta n$ determines the optimal cooling limit. As defined above, $\Delta n=(2\pi\dot{E})/(\hbar\omega^2)$ is determined both by the laser parameters (beam waist, wavelength, power) and the particle material properties (radius, dielectric constant, mass density). It is remarkable that all these parameters can group into one single variable $\Delta n$.

In the rest of this subsection, we explore the trend of force feedback cooling as the parameter set is tuned. On the one hand, we numerically demonstrate that the parameter set \{$\Delta n$,$\eta$,$\tilde{\gamma}$\} indeed controls the dynamics (the cooling limit stays the same as long as the parameter set is fixed, no matter what material, beam waist or laser power are used). According the Eqs. (\ref{5e2}) and (\ref{5e44}), $\Delta n$ is shown to be
\begin{equation}
\Delta n=2\pi\frac{\dot{E}}{\hbar \omega^2} =  \frac{\pi }{3}\frac{\epsilon-1}{\epsilon+2}R^3w_0^2 k_0^5,
\end{equation}
where $k_0=2\pi/\lambda$ is the incoming wave vector and $w_0$ is the beam waist. One finds that $\Delta n$ is independent of the laser power, which means changing laser power has no effect on the steady state occupation number if all other parameters are fixed. This is shown in Fig. (\ref{5fg3}) with $\eta=0.1$ (the results with other $\eta$ are similar), where the steady state occupation numbers in terms of the scaled feedback strength are exactly the same for the case with different trapping laser powers. Further, as we change $\Delta n$ (by tuning $w_0$, $\alpha$, or $k_0$), the steady state occupation number with respect to $\tilde{\gamma}$ varies, as shown by the curves with $\Delta n=0.033$ and $\Delta n=0.076$. It is worth noting that Fig. (\ref{5fg3}) also indicates the good agreement between semi-classical and quantum calculation as shown before. 

\begin{figure}
\includegraphics[width=8.0cm,height=5.5cm]{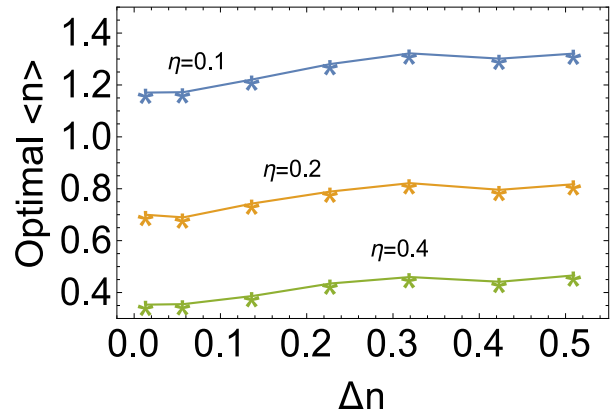}
\caption{{(Color online)} The optimal cooling limit in terms of the parameter $\Delta n$ (by tuning the laser beam waist). The blue, yellow and green curves correspond to the cases with measurement efficiency $\eta$ fixed at $(0.1,0.2,0.4)$. The data is based on semi-classical calculations. \label{5fg4}}
\end{figure}

The optimal cooling limit only depends on $\Delta n$ and $\eta$. As shown in Fig. (\ref{5fg2}), a higher measurement efficiency $\eta$ will lead to a lower steady state occupation number. However, it is not obvious how the optimal cooling limit depends on $\Delta n$. Figure (\ref{5fg3}) seems to show that the optimal cooling limits don't vary strongly with different $\Delta n$. To understand the role of $\Delta n$, we calculate the optimal cooling limit for several measurement efficiencies as a function of $\Delta n$. The result is shown in Fig. (\ref{5fg4}). The optimal cooling limit weakly depends on the parameter $\Delta n$. For measurement efficiencies $\eta=(0.1,0.2,0.4)$, the optimal cooling limit weakly increases as we increase the parameter $\Delta n$. The fact that $\Delta n$ has little effect on the optimal cooling limit essentially means the measurement efficiency $\eta$ is the most important parameter affecting the optimal cooling limit using force feedback. For reference, we list the optimal cooling limit with varied parameters in Tab. \ref{tab2} (shown in Appendix \ref{a2}).

%-------
%-------

\subsection{Cooling by parametric feedback}

In this subsection, we present the simulation results of the parametric feedback cooling by solving the SSE and the semi-classical equations. The SSE for parametric feedback cooling is already introduced in section \ref{s2}. The semi-classical parametric feedback cooling is discussed in Ref. \cite{SNDR}. For demonstration, we list below the semi-classical equations for parametric feedback cooling
\begin{equation}
\begin{split}
m\frac{d^2x}{dt^2}&=-m\omega^2(1+\chi x_m\dot{x}_m)x,\\
x_i&=x+dW\cdot\Delta x,\\
p(t+dt)&=p(t)+dW\cdot\Delta p,
\end{split}
\end{equation}  
where $\chi$ is the parametric feedback strength and the other quantities are the same as those in the force feedback equations. The semi-classical equation for parametric feedback cooling can also be scaled and the dynamics was shown to be dependent on the parameter set: the measurement efficiency $\eta$, the scaled parametric feedback strength $\tilde{\chi} =\hbar \chi/(2 m)$ and the $\Delta n$ (Eq. (\ref{5e13})). One can refer to Ref. \cite{SNDR} for a detailed discussion. Similar to solving the force feedback cooling, the semi-classical equations of motion for parametric feedback cooling are numerically solved using a fourth-order Runge-Kutta algorithm. The momentum kick is added to the nanoparticle at each time step. The SSE is solved in a harmonic eigen-basis and details are presented in the appendix \ref{a1}. All simulations are repeated over one thousand times and data is collected by averaging over the different runs. 

\begin{figure}
\includegraphics[width=8.0cm,height=5.5cm]{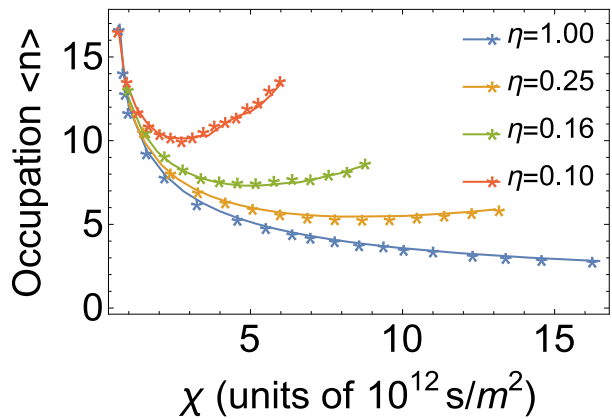}
\caption{{(Color online)} The steady state occupation in terms of the parametric feedback strength for diamond with parameters given in Tab. \ref{tab1}. The solid line is results from semi-classical calculation while the asterisks correspond to the results from SSE. The blue, yellow, green, and red color correspond the calculations with measurement efficiencies $\eta=(1.0,0.25,0.16,0.1)$ respectively. \label{5fg5}}
\end{figure}

\begin{figure}
\centering
\subfigure
{
\put(22,-18){(a)}
\begin{minipage}[]{0.22\textwidth}
\includegraphics[width=4.2cm,height=3.2cm]{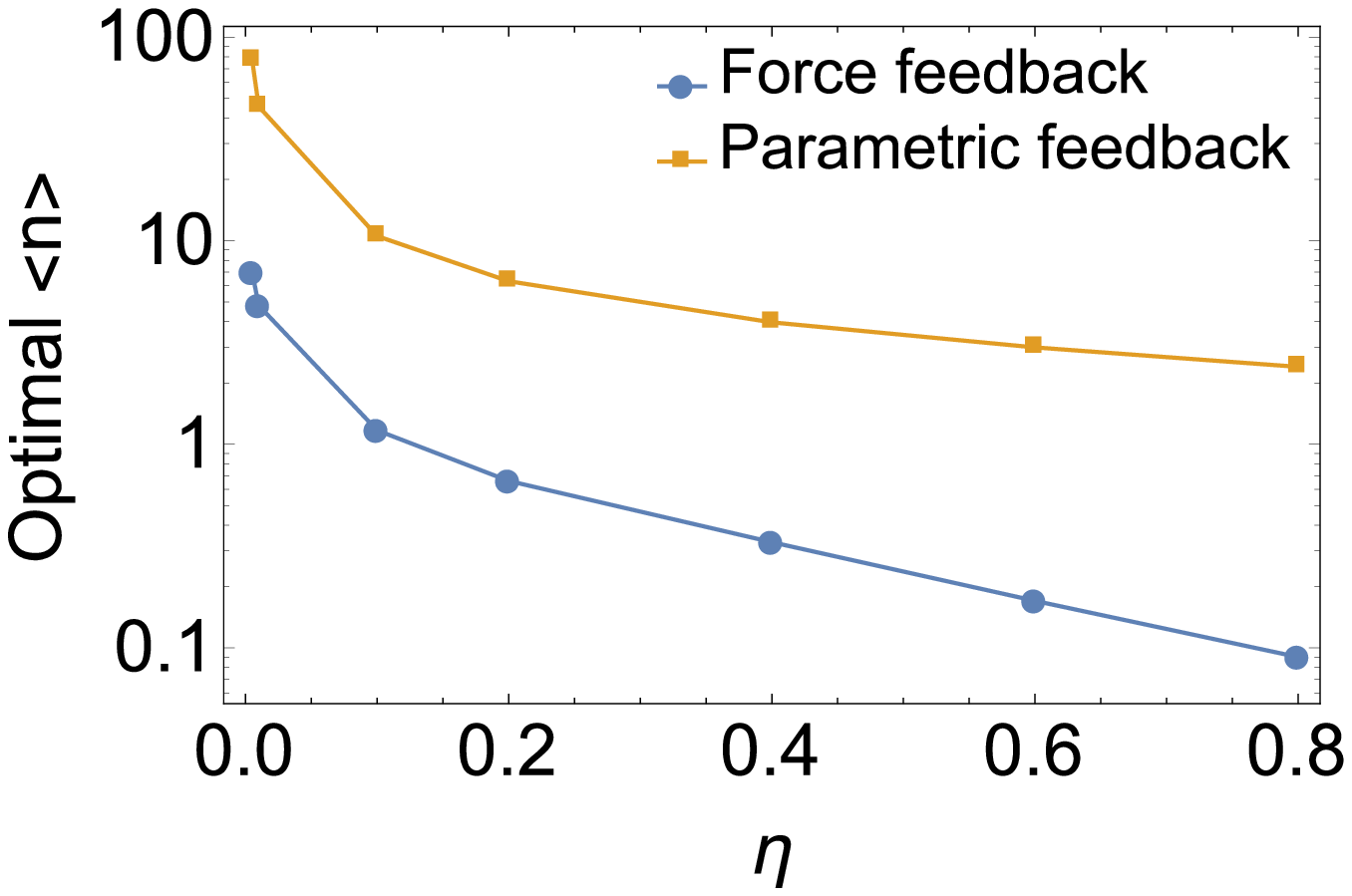}
\end{minipage}
\label{5fg6a}
}
\subfigure
{
\put(22,-18){(b)}
\begin{minipage}[]{0.22\textwidth}
\includegraphics[width=4.2cm,height=3.2cm]{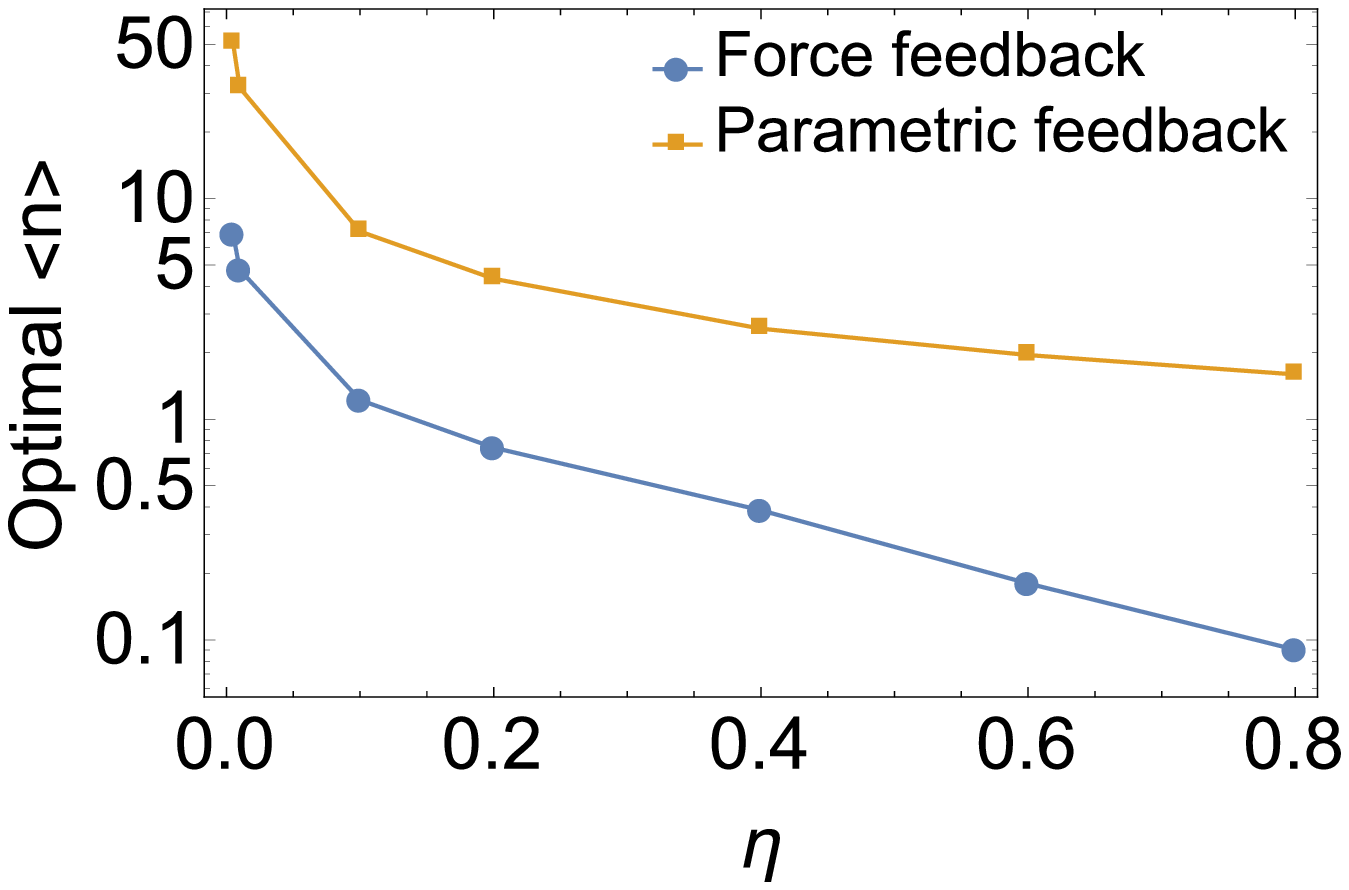}
\end{minipage}
\label{5fg6b}
}
\caption{ {(Color online)} The optimal cooling limit with respect to the measurement efficiency. The blue lines connect the results from force feedback cooling, while the yellow lines connect the results from parametric feedback. The $y$ axes are given in log scales. The data is obtained from solving semi-classical equations. \textbf{(a)} $\Delta n=0.0142$. \textbf{(b)} $\Delta n=0.1372$.  \label{5fg6} } 
\end{figure}

Figure (\ref{5fg5}) gives the steady state occupation number in terms of parametric feedback strength, which has a similar structure to the force feedback cooling. First, the quantum and semi-classical results also match very well. Second, for different measurement efficiencies, there is also a minimal point (the optimal cooling limit) which can be reached when the parametric feedback strength is tuned. Comparing Fig. (\ref{5fg2}) with Fig. (\ref{5fg5}), the optimal cooling limit from the parametric feedback cooling is higher than that by force feedback. This indicates that ground state cooling by force feedback may be favored over parametric feedback. To clearly see that, we perform a calculation and collect the optimal cooling limit with respect to the measurement efficiency for the two cooling schemes. Figure (\ref{5fg6}) gives the result, which shows a much lower occupation number can be reached using the force feedback when the same measurement efficiency is used.

%-------
%--------

Similar to the force feedback, the parametric feedback cooling also only depends on the parameter set \{$\eta$,$\tilde{\chi}$,$\Delta n$\}. First, as shown in Fig. (\ref{5fg5}), a lower optimal cooling limit can be obtained if one increases the measurement efficiency, which is the same as the force feedback cooling. However, the dependence of the optimal cooling limit on $\Delta n$ is quite different. As one increases $\Delta n$, the optimal cooling limit is observed to decrease significantly. The detailed discussion can be found in Ref. \cite{SNDR}. Thus, unlike the force feedback cooling, one can efficiently tune both $\eta$ and $\Delta n$ so as to parametrically cool the levitated nanoparticle. For reference, we list the optimal cooling limit from parametric feedback with varied parameters in Tab. \ref{tab3} (shown in Appendix \ref{a2}).

%------------------

\begin{figure}
\includegraphics[width=8.0cm,height=5.5cm]{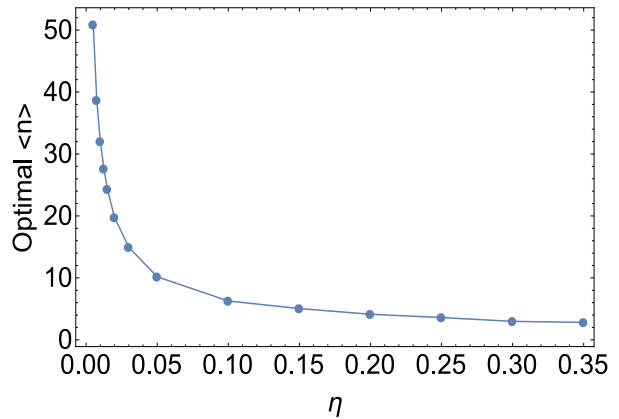}
\caption{The simulation of an experiment \cite{MKCO} of parametric cooling of a fused silica, with $\Delta n\simeq0.9$. The plot gives the optimal cooling limit in terms of the measurement efficiency $\eta$. The measurement efficiency is scanned from $\eta=0.005$ to $\eta=0.35$. The occupation number gets below $\braket{n}=20$ when the measurement efficiency $\eta\geq0.015$. \label{5fg7}}
\end{figure}

In Ref. \cite{DMOP}, $\braket{n}=63$ was reached using parametric cooling. In a recent experiment  \cite{MKCO}, $\braket{n}=21$ was reached due to a better detection efficiency. In the experiment, a fused silica nano-sphere with radius about $R=50$ nm is trapped in a polarized laser beam with wavelength $\lambda=1064$ nm. The silica has a dielectric constant $\epsilon=2.1$ and a mass of about $m=1.13\times10^{-18}$ kg. The oscillation frequency in one transverse degree of freedom is measured to be $\omega=2\pi\times143$ kHz, which corresponds to an effective numerical aperture NA$\simeq0.5$. The shot noise in this degree of freedom is measured close to $\braket{\dot{n}}\simeq 21$ kHz \cite{DMOP}. Combining the above the parameters, we arrive at $\Delta n\simeq0.9$. With these parameters, we simulate the parametric feedback cooling by scanning the measurement efficiency. The result is shown in Fig. (\ref{5fg7}), where an occupation number lower than $20$ can be reached if the measurement efficiency is more than $\eta=0.015$, and lower occupation number can be reached when the measurement efficiency increases.

%-------
%--------

\section{Conclusion}\label{s4}

In summary, we have extended the semi-classical calculation of the feedback cooling of a laser levitated nanoparticle in the shot-noise-dominant regime to the quantum domain. Using the theory of continuous quantum measurement, the measured particle position can be obtained continuously and the system state evolves stochastically due to the measurement back action. Cooling is achieved by feeding back the measured information (force and parametric feedback). Similar to the results from a model of semi-classical feedback scheme, the quantum cooling only depends on the feedback strength, $\Delta n$ (the change of occupation number in one vibrational period) and the measurement efficiency. The minimum occupation number only depends on $\Delta n$ and the measurement efficiency. The match between quantum and semi-classical results suggests that one can perform the much faster and more intuitive semi-classical calculation when analyzing the cooling of a levitated nanoparticle. The comparison between parametric feedback and force feedback cooling reveals that the force feedback cooling scheme is easier to reach the ground state. 

\section{Acknowledgement}

This work was supported by the National Science Foundation under Grant No.1404419-PHY and No.1555035-PHY.

\begin{appendix}

\section{The numerical method for solving the SSE and the SME} \label{a1}

This appendix introduces the numerical schemes used in the main text. The SSE (Eq. \ref{5e4}) is used in the calculations of the force feedback and parametric feedback cooling. The Hamiltonian is given by
\begin{equation}
\bm{H}=\bm{H}_0+\bm{H}_{1,2},
\end{equation}
where $\bm{H}_0=\hbar\omega(\bm{a}^\dagger\bm{a}+\frac{1}{2})$, $\bm{H}_1=\gamma p_m\bm{x}$ denotes the force feedback and $\bm{H}_2=\frac{\chi}{2}m\omega^2x_m\dot{x}_m\bm{x}^2$ is the parametric feedback. At each time step, the Hamiltonian $\bm{H}$ is modified according to the measured position $x_m$ and Eq. (\ref{5e9}) is used in evaluating $\bm{H}_1$ and $\bm{H}_2$. To get a better estimate of the particle position, $x_m$ at time t is calculated by taking the weighted time average of the directly measured positions in the earlier time $x_i(t^\prime)$, 
\begin{equation}
x_m(t)=\frac{1}{\mu}\int_{-\infty}^t x_i(t^\prime)e^{-\frac{(t-t^\prime)}{\mu}}dt^\prime,
\end{equation}
where $\mu\ll T$ and $T=2\pi/\omega$ is the oscillation period. In our calculation, we take $\mu=T/20$. The wave function $\ket{\psi(t)}$ is represented in the eigen-basis of the operator $\bm{H}_0$. The initial state is chosen to be Gaussian, which is sensible because any other initial state would evolve rapidly into Gaussians under the continuous monitoring \cite{FCOQ,LQTA,CSVD,EOQS}. The numerical propagation of Eq. (\ref{5e4}) is split into two parts. The first part is the unitary evolution $d\ket{\psi}=-\frac{i}{\hbar}\bm{H}dt\ket{\psi}$ which is solved by the well known Crank-Nicolson method \cite{NRIC}. For the second part, the increment $d\ket{\psi}=(-\kappa(\bm{x}-\braket{\bm{x}})^2dt
+\sqrt{2\kappa}(\bm{x}-\braket{\bm{x}})\sqrt{dt}dW)\ket{\psi}$ is directly calculated in the eigen-basis, and the random number $dW$ is generated and used at each time step. The wave function in the next time step is obtained by renormalizing the sum from the first and the second part of the propagation. Each calculation is repeated more than one thousand times and data is collected by averaging over them. The convergence is checked by changing the time step size as well as the number of eigenstates used in the simulation. 

The SME (Eq. \ref{5e3}) is used in the force feedback cooling calculation. The density operator is also represented in the eigen-basis of Hamiltonian $\bm{H}_0$. Equation (\ref{5e3}) is numerically solved using a second order Runge-Kutta algorithm. At each time step, a random number is generated and used, and the measured position $x_m$ is used to get a feedback signal $p_m$, such that a modified Hamiltonian $\bm{H}$ is obtained. The simulation is performed many times and data is collected by averaging over a thousand trajectories. The convergence is also checked by changing the time step size and the number of eigenstates. The SME is basically equivalent to the SSE, so the results are expected to match when the same values of the parameters are used.

\section{The data for optimal cooling limit} \label{a2}

For reference, this appendix lists the data of optimal cooling limit with respect to the parameter set from both the force feedback and the parametric feedback cooling.
\begin{table}[htp]
\caption{This table gives the optimal cooling limit from the force feedback cooling scheme in terms of the parameters $\eta$ and $\Delta n$. Each data point is obtained by scanning the feedback strength. The data roughly follows the formula $\braket{n}=\frac{0.48}{\sqrt{\eta}}-\eta+\frac{0.15}{\eta^{{1}/{3}}}\Delta n-0.01\Delta n$. } \label{tab2}
\begin{center}
\begin{tabular}{|c|c|c|c|c|c|c|c|c|c|}
\hline
\diagbox{$\eta$}{$\Delta n$} & $0.01$ & $0.05$  & $0.10$  & $0.15$  & $0.20$  & $0.25$ & $0.30$ \\
\hline
0.005 & $6.71$ & $6.80$  & $6.85$   &   $6.92$   &  $6.93$   &   $6.95$ & $6.99$  \\
\hline
0.01 & $4.62$ & $4.69$  & $4.73$   &   $4.76$   &  $4.81$   &   $4.86$ & $4.90$  \\
\hline
0.05 & $1.82$ & $1.86$  & $1.88$   &   $1.90$   &  $1.92$   &   $1.94$ & $2.00$   \\
\hline
0.1 & $1.13$ & $1.17$  & $1.18$   &   $1.22$   &  $1.27$   &   $1.30$ & $1.31$   \\
\hline
0.2 & $0.68$ & $0.70$  & $0.71$   &   $0.74$   &  $0.76$   &   $0.80$ & $0.79$  \\
\hline
0.4 & $0.34$ &  $0.35$  & $0.36$   &   $0.39$   &  $0.42$   &   $0.44$ & $0.45$  \\
\hline
\end{tabular}
\end{center}
\label{default}
\end{table}

\begin{table}[htp]
\caption{This table gives the optimal cooling limit from the parametric feedback cooling scheme in terms of the parameters $\eta$ and $\Delta n$. The data stops at $\Delta n=0.2$ since our calculation becomes unstable for bigger values of $\Delta n$. Each data point is obtained by scanning the feedback strength. }\label{tab3}
\begin{center}
\begin{tabular}{|c|c|c|c|c|c|c|c|c|c|}
\hline
\diagbox{$\eta$}{$\Delta n$}  & $0.01$  & $0.05$  & $0.10$  & $0.15$  & $0.20$  \\
\hline
0.005 & $129$  & $65.9$   &   $56.0$   &  $48.0$   &   $43.9$  \\
\hline
0.01 & $76.6$  & $42.1$   &   $36.0$   &  $31.0$   &   $28.9$  \\
\hline
0.05 & $35.0$  & $19.0$   &   $16.3$   &  $14.1$   &   $12.2$  \\
\hline
0.1 & $15.1$  & $9.05$   &   $7.81$   &  $6.89$   &   $6.30$   \\
\hline
0.2 & $9.40$  & $5.72$   &   $4.71$   &  $4.19$   &   $3.86$   \\
\hline
0.4 &  $6.27 $  & $3.53 $   &   $2.89$   &  $2.49$   &   $2.43$  \\
\hline
\end{tabular}
\end{center}
\label{default}
\end{table}

\end{appendix}

\bibliographystyle{ieeetr}
\bibliography{all.bib}

\end{document}